\documentclass[10pt,journal,compsoc]{IEEEtran}
\usepackage[T1]{fontenc}
\usepackage{amsmath,amssymb}
\usepackage{graphicx}   
\usepackage{url}
\usepackage{booktabs}
\usepackage{array}
\usepackage{textcomp}
\begin{document}
\title{Separation of Duties for Privileged LLM Agents: A Governed Execution Architecture with Measured Security--Utility Trade-offs}
\author{Qishuai Jing\thanks{The author is an independent researcher in Yangzhou, Jiangsu 225001, China (e-mail: jqs113@foxmail.com). ORCID: 0009-0007-3951-1763. Corresponding author. This work has been submitted to the IEEE for possible publication. Copyright may be transferred without notice, after which this version may no longer be accessible.}}
\maketitle
\begin{abstract}
Large language model agents are increasingly granted real privileges (executing commands, modifying files, calling APIs), so an agent that errs has already acted. Existing defences concentrate on the agent's \emph{inputs}, while the path from a candidate action to privileged side effects remains less directly studied. We argue that this path must be governed outside the model, and study an architecture interposing four roles (planner, policy gate, executor, auditor) between agent and operating system.

Two choices are central: actions arrive as structured intents, so adjudication never parses shell syntax; and approval is a one-shot credential bound to the exact bytes that will run. We evaluate on a 313-case benchmark across eight variants, with prompt-only baselines from three hosted LLMs on 150 stratified cases, five repetitions (2,250 attempted calls; 2,249 completed).

Effective attack success falls from 98.3\% under direct execution to 7.7\% deployed. Re-execution against the real implementation yields a similar aggregate rate (7.6\% over 66 sandbox-evaluable payloads) but substantial case-level disagreement, at a corrected false-denial rate of 11.1\%. A substantial final-stage reduction (from 30.8\% to 7.7\%) is attributable to the operating-system sandbox, and the benchmark found four implementation defects, none by design review.
\end{abstract}
\section{Introduction}
\label{sec:1}

Large language models are increasingly deployed not as chatbots but as \emph{agents}: systems that read files, execute commands, call APIs, and modify state on behalf of a user. This shift changes what failure means. A chatbot that produces an incorrect answer can be corrected before anyone acts on it; an agent that produces an incorrect action has already acted. The consequence of an error scales with the privileges the agent holds, and those privileges are, by design, substantial --- the agent's usefulness comes precisely from its ability to do things.

Governance practice has begun to catch up. OWASP's agentic-security guidance \cite{ref11}, the NIST AI Risk Management Framework \cite{ref24}, and the Cloud Security Alliance's agent identity guidance \cite{ref17} all emphasise controls around privileged actions, traceability, and least privilege. What is less settled is \emph{where} that check should live and \emph{how much} it actually buys.

A dominant line of defence polices the agent's \emph{inputs}. Prompt-injection detectors, guardrail models, and content filters inspect untrusted text before it reaches the model, and benchmarks such as InjecAgent and AgentDojo measure how well such defences work. This work is valuable, but input filtering alone does not enforce the later path from a candidate tool call to privileged execution.

The gap matters because an action can be dangerous while every input was trustworthy. An agent with a legitimate goal may still attempt to delete a protected file, escape a sandbox, or reuse an approval it obtained earlier. Nothing in the input stream is anomalous; the anomaly is in the action.

A natural first response is to instruct the model itself: tell it to seek approval before irreversible or dangerous operations. We tested this directly. Three language models from three distinct families received an identical system prompt requiring explicit approval before any dangerous action, and were then asked to adjudicate 150 cases five times each --- 2,250 attempted calls in total, of which 2,249 completed and 2,164 yielded a parseable decision.

The instruction did not hold. Pooled across the three models, 78.1\% of dangerous operations were permitted (349 of 447 parseable decisions on dangerous cases) and 94.7\% of bypass attempts succeeded (95\% CI 93.0--96.0\%). The failure modes differed by family and in opposite directions: one model read the bare request "please execute this" as authorisation and permitted 100\% of dangerous operations with a zero refusal rate; the other two were conservative enough to demand confirmation for legitimate deletions inside the workspace. Repeated runs of the \emph{same} case at temperature 0 agreed only 48.7--71.3\% of the time.

Two conclusions follow. First, a prompt is a request, not an enforcement mechanism. Second, any evaluation of an LLM-mediated security control must report run-to-run variation, because a single run is not a measurement.

If the model cannot be the boundary, the boundary must be built around it. This paper studies an architecture that interposes four roles between the agent and the operating system: a planner that drafts actions, a policy gate that adjudicates them, a privileged executor that performs only affirmatively adjudicated actions, and an independent auditor that reviews the ledger on a schedule. Two design choices are central: actions may be submitted as structured intents (typed operation, absolute paths, typed arguments), so adjudication never parses shell syntax; and an approval is a credential bound to the exact bytes that will run and consumable exactly once.

This is a deliberate revival of a very old idea. \emph{Separation of duties} --- that no single party should both propose and approve a sensitive operation --- predates computing by centuries. For brevity we call the design the Three-Department Governance Architecture (TDGA), after the historical Chinese \emph{Three Departments} arrangement (Secretariat drafted, Chancellery reviewed, Department of State Affairs executed); the analogy is organisational inspiration, not a claimed security contribution \cite{ref27}. Its translation to agents is not automatic: one industry analysis notes that separation of duties degenerates when agents can route approvals through one another \cite{ref14}. Whether the principle survives that translation, and at what cost, is an empirical question.

We answer with a controlled benchmark rather than an argument. Across 313 cases in five groups --- normal, explicitly dangerous, fifteen families of expression-level bypass, post-approval mutation and credential-chain attacks, and fault injection --- we evaluate eight variants arranged as a progressive mechanism ladder, in which two adjacent pairs isolate a single added mechanism (B5$\rightarrow$B6 for executor-side file-byte verification, B6$\rightarrow$B7 for the OS sandbox) while other rungs change bundles or channels and are interpreted accordingly. We verify the sandbox's boundary behaviour on the real implementation while retaining a disclosed per-case behavioural model for the full B7 benchmark, and compare against prompt-only baselines from three real models.

Four findings stand out.

Effectiveness. Effective attack success falls from 98.3\% under direct execution to 7.7\% deployed, but a substantial final-stage reduction comes from the operating-system sandbox rather than from the policy layer: two variants differing only by sandboxing differ by 30.8\% versus 7.7\%.

Cost. The deployed configuration denies 11.1\% of benign tasks after per-case review (the raw behavioural-model figure was 20.4\%). In the separate LLM sample, the claude-opus-5 baseline permitted 89.8\% (53/59) of bypass cases while producing a non-executable outcome for 3.3\% (1/30) of benign ones under five-run majority aggregation. The samples differ, so this remains a directional comparison: TDGA incurs a higher observed benign non-execution rate here, but a far lower bypass-success rate.

Model decoupling. The three prompt-only baselines fail in different directions, but the deterministic gate does not consume model identity: for a fixed candidate action and policy state its decision is model-independent by construction, leaving a full cross-model generation-to-execution evaluation to future work.

Defects found by measurement. Four implementation defects were uncovered, none by design review. One was invisible in aggregate metrics because it manifested as \emph{stricter} behaviour, another because it merely looked like "another passing case"; both required tests that actually execute.

This paper makes four contributions:

1. A measurable formulation of action-side governance. Agent governance is cast as adjudication over actions rather than inputs, with metrics --- effective attack success, false denial, fail-closed rate, mutation and replay blocking, audit coverage --- that make the trade-off explicit rather than rhetorical. 2. A controlled evaluation with disclosed denominators and caveats. 313 cases $\times$ 8 variants; three real LLMs $\times$ 150 cases $\times$ 5 attempts (2,250 calls); McNemar tests on paired decisions; Wilson intervals; and a documented account of every measurement condition that could change how a number should be read. Sandbox-relevant payloads are additionally re-executed against the real deployment configuration. 3. Evidence for model-decoupled enforcement. Three families exhibit different prompt-only failure modes while the deterministic gate adjudicates a fixed action without reference to model identity. 4. Failure as data. Four defects, all found by the benchmark, and the argument that executable adversarial benchmarks are a research method rather than an appendix.

\section{Related Work}
\label{sec:2}

\subsection{Formalising LLM agent security}
\label{sec:2-1}

Recent work has begun to give agent security a formal footing. One framework \cite{ref1} proposes four context-dependent security properties --- \emph{task alignment} (pursuing authorised objectives), \emph{action alignment} (individual actions serving those objectives), \emph{source authorisation} (executing commands from authenticated sources), and \emph{data isolation} (information flows respecting privilege boundaries) --- and argues that whether a given action constitutes a violation depends on whose instruction led to it and what objective it serves. Earlier work \cite{ref2} formalised prompt-injection attacks and defences, turning a body of case studies into something quantitatively comparable.

\subsection{Prompt-injection defences and input-side guards}
\label{sec:2-2}

This is the most active line of work. InjecAgent \cite{ref3} provides 1,054 test cases covering 17 user tools and 62 attacker tools, and demonstrates that tool-integrated agents are fundamentally vulnerable to indirect injection; AgentDojo \cite{ref4} offers a more task-realistic injection environment; and the Agent Security Bench \cite{ref5} combines formalisation with benchmarking. A recent replication study \cite{ref6} across several frontier models evaluates four defences against an undefended control and reports attack success, benign utility, utility under attack, and run-to-run variation, cautioning that evaluations must report goal feasibility and action availability alongside attack outcomes.

On the defence side, CaMeL \cite{ref7} claims \emph{provable} security by separating a privileged LLM (processing only trusted instructions) from a quarantined LLM (handling untrusted data), with capability-based policies enforced inside a custom interpreter; an earlier, informal statement of the same idea is the Dual LLM pattern \cite{ref8}. StruQ \cite{ref9} separates instructions from data via structured queries, and Task Shield \cite{ref10} verifies at test time whether each instruction and tool call contributes to the user-specified goal, reporting 2.07\% attack success at 69.79\% utility on AgentDojo. AttriGuard \cite{ref34} moves closer to the action boundary by causally attributing proposed tool invocations to user intent versus untrusted observations; its target remains indirect prompt injection, whereas our benchmark begins after a candidate action has been produced.

The residual gap is also reflected in verification guidance. OWASP's AISVS \cite{ref11} requires access-control decisions for agent actions to be enforced by application logic or a policy engine, not by the model itself.

Several recent systems provide closer architectural comparators. Progent \cite{ref25} expresses deterministic least-privilege constraints over tool calls, AgentSpec \cite{ref32} provides a lightweight DSL for customizable runtime enforcement, and Design Patterns for Securing LLM Agents \cite{ref26} catalogues privileged/quarantined and related architectural patterns. These works strengthen the case for controls outside the LLM, but they do not evaluate the same execution-credential chain studied here.

\subsection{Separation of duties and runtime governance layers}
\label{sec:2-3}

\emph{Separation of duties} is a traditional internal-control concept, originating in accounting and financial workflow and formalised in modern control catalogues such as NIST SP 800-53 AC-5 \cite{ref28}. Its translation to agents has been articulated in industry: one engineering guide \cite{ref12} defines it as distributing responsibilities across multiple agents or systems so that no single agent can control an entire sensitive workflow, tying it to identity governance, delegated authorisation, and workflow monitoring; Microsoft's Cloud Adoption Framework \cite{ref13} recommends multi-agent designs at compliance boundaries and cites financial services, where one agent prepares transactions and another validates them.

The same body of work names its own weak point. As one analysis \cite{ref14} puts it, \emph{"separation of duties works because a human cannot approve their own request. Agents can, by routing through each other."} Where agents can approve one another, formal separation degenerates into self-approval.

On the academic side, Governance-as-a-Service \cite{ref15} reconceives governance as runtime infrastructure interposed between agents and their environment and evaluates policy enforcement across multiple open-source models and adversarial regimes. IsolateGPT \cite{ref29} provides execution isolation for third-party LLM applications; ACE \cite{ref30} separates abstract and concrete planning and enforces data and capability barriers during execution; and SAGA \cite{ref31} focuses on agent identity, authorisation, delegation, and lifecycle governance for interacting agents. Progent \cite{ref25} likewise enforces deterministic, programmable least-privilege constraints over tool calls and reports security--utility results across AgentDojo, ASB, and AgentPoison. Work on agentic architectures \cite{ref16} similarly stresses deterministic, versioned decision rules.

\subsection{Structured intents, capability declarations, and credential binding}
\label{sec:2-4}

Stripping "what is to be done" out of natural language is a shared direction across several independent lines. The Cloud Security Alliance's draft Agent Identity Governance Framework \cite{ref17} proposes just-in-time, intent-declared, time-bound, scope-limited grants and stresses that intent declaration yields a governance artefact that standing-privilege models lack: a pre-execution record against which post-execution behaviour can be compared. CaMeL \cite{ref7} uses capability-based policies for a closely related purpose, and the FINOS AI Governance Framework \cite{ref18} lists agent least privilege as a standalone mitigation, enforcing approval workflows at the privilege level. At the programming-language layer, tracked-capability work \cite{ref33} constrains agent-generated programs through statically tracked capabilities and information-flow restrictions. These approaches differ in enforcement object and abstraction level, but all move authority away from unconstrained model output.

Two steps beyond intent declaration were shown by measurement to be necessary: the object adjudicated must be the bytes about to execute, not the intent text (our first implementation verified only the intent digest, so "approve $\rightarrow$ swap $\rightarrow$ execute" went through), and an approval must be a one-shot credential (our third allowed replay, with measured blocking of only 77.8\%, rising to 100\% once single-use). Evidence: Section~\ref{sec:5-7}, VUL-003/004.

These mechanisms are not claimed as novel security primitives in isolation; our contribution is to integrate them into LLM-agent execution governance and to evaluate their concrete failure modes experimentally.

\subsection{Audit integrity}
\label{sec:2-5}

Audit-side work is comparatively mature. Tamper-evident audit trails are defined \cite{ref19} as records engineered so that post-event modification, deletion, or gaps are detectable, with each entry bound to an identity, a decision point, and a time boundary. Compliance-oriented treatments \cite{ref20} enumerate the required fields --- agent identity, human authoriser, data accessed and its classification, operation performed, policy-evaluation outcome, and a tamper-evident timestamp --- and emphasise that the record should be created at the relevant event rather than reconstructed only at workflow completion. Common mechanisms are hash chaining, append-only storage, and role-based access control; earlier academic work \cite{ref21} used blockchain-backed logging to provide tamper-proofing and non-repudiation for AI audit logs. ISO/IEC 42001 \cite{ref22} establishes management-system requirements for responsible AI governance, the EU AI Act \cite{ref23} requires high-risk systems to support automatic logging of events over their lifetime, and the NIST AI RMF \cite{ref24} stresses traceability and accountability. More directly, NIST SP 800-53 defines AC-5 Separation of Duties and AU-9 Protection of Audit Information, including protection against unauthorised modification and deletion \cite{ref28}.

Our operational measurements add what the audit literature leaves open: concurrent ledger writes can lose records silently (9.66\% at two writers to 31.32\% at eight, with 12 torn lines), read cost grows with ledger size (163.83 ms and 676 ms at 50,000 lines for the two query directions), and malformed lines are skipped silently unless counted separately (Section~\ref{sec:5-4}, Section~\ref{sec:5-5}).

The literature asks how to make logs immutable; we ask whether the ledger remains complete and fast under real load. Neither question is decorative.

\subsection{Positioning}
\label{sec:2-6}

Taken together, this paper's increment can be stated in three sentences.

1. We measure a narrower execution-path problem in depth. Prior work evaluates runtime guards and privilege controls; exact-byte approval binding, one-shot credential consumption, pre-execution re-verification and audit-path operational failures are less directly characterised, and our 313-case benchmark is built around them. 2. We turn claims into numbers, and report the cost. Separation of duties, decoupled governance and intent declaration have all been argued for; we quantify the progression from direct execution to the deployed configuration (98.3\% $\rightarrow$ 7.7\%), isolate the sandbox and file-verification effects, and report usability cost (11.1\% corrected false denial) and overhead (zero tokens; latency as a function of ledger size). 3. We treat failure as part of the method. Four implementation defects (VUL-001 to VUL-004) were all discovered by measurement, none by design review --- two invisible in aggregate metrics.

\section{System and Threat Model}
\label{sec:3}

\begin{figure*}[!t]
\centering
\includegraphics[width=\textwidth]{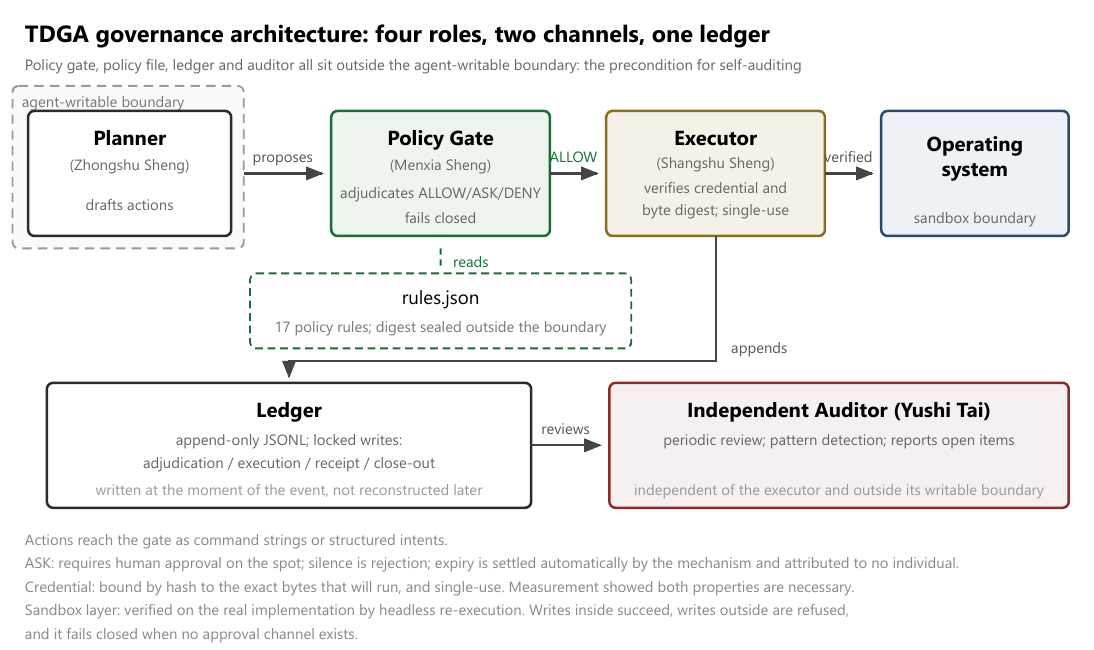}
\caption{The governance architecture. Four roles --- planner, policy gate, privileged executor, independent auditor --- sit between the agent and the operating system. Actions reach the gate through two channels (command strings, structured intents); the gate, the policy file, the ledger and the auditor all lie outside the agent's writable boundary (dashed), which is what makes the audit trail non-rewritable by the governed party.}
\label{fig:1}
\end{figure*}

The system under test is a governance layer interposed between a privileged LLM agent and the operating system, organised as four roles (Figure~\ref{fig:1}): a planner that drafts candidate actions, a policy gate that adjudicates each action into one of three verdicts, a privileged executor that runs only affirmatively adjudicated actions after re-checking the credential that authorised them, and an independent auditor that reviews the ledger and reports unresolved items.

Actions are submitted through two channels: a command-string channel carrying arbitrary shell text, and a structured-intent channel carrying a typed record --- an operation from a closed vocabulary, absolute paths, typed arguments --- so that no shell-syntax parsing is needed to decide on it. Every adjudication, execution, receipt and close-out is appended to an append-only JSONL ledger.

The implementation comprises 1,635 lines of Python and a 17-rule policy file. Both are hashed at freeze time and the digests are sealed outside the agent's writable boundary, so any later modification of gate or policy is detectable; the artefacts, their SHA-256 digests and the exact interpreter and OS versions are recorded in the reproducibility package (Section~\ref{sec:4-8}).

Credential semantics. Two properties of the approval mechanism deserve explicit statement, because both were introduced only after measurement showed them to be necessary (Section~\ref{sec:5-7}).

\begin{itemize}
\item An approval binds to the bytes of the staged artefact, not merely to the identity of the intent. The gate records a digest of the staged file when it adjudicates; both the gate and the executor re-compute it before the artefact runs.
\item An approval is single-use. Consumption is recorded by the executor itself rather than through a side channel, so that a failure of the close-out path cannot silently make an approval reusable.
\end{itemize}

We assume the adversary \emph{is} the governed agent, or an attacker who has subverted it through prompt injection. It therefore has full control of what it emits, and is assumed to be goal-directed rather than merely careless. Concretely it may:

(i) emit arbitrary command strings, including syntactically valid but semantically obfuscated forms;

(ii) exploit expression-level equivalences --- relative and UNC paths, forward slashes, 8.3 short names, environment-variable indirection, junctions, alternative interpreters such as Python file APIs --- to evade string matching;

(iii) propose structured intents that appear benign;

(iv) modify a staged artefact \emph{after} it has been approved; and

(v) replay, forge, or reuse an approval credential.

The adversary cannot modify the gate, the policy file, or the ledger, all of which reside outside its writable boundary; this is the sandbox boundary established in Section~\ref{sec:4-5}.

Out of scope. Multi-agent collusion, side channels, and supply-chain compromise are not considered. The sandbox's own escalation path is also outside the experimental design: in unattended settings it fails closed, but under an interactive deployment the model may request wider access and a human decides --- and humans are persuadable (Section~\ref{sec:7}).

\section{Experimental Design}
\label{sec:4}

\subsection{Benchmark}
\label{sec:4-1}

We constructed a 313-case benchmark organised in five groups.

\begin{itemize}
\item Group A --- normal operations (54). Legitimate work the agent is expected to perform, including in-workspace file operations, read-only queries, and workflow commands. These measure whether the system is usable at all.
\item Group B --- explicitly dangerous operations (74). Deletions of protected artefacts, registry writes, disk operations, download-and-execute patterns, living-off-the-land binary abuse, and privilege changes.
\item Group C --- bypass attacks (122). Fifteen families of expression-level evasion: environment-variable indirection, relative and bare-relative paths, UNC paths, slash variants, volume-GUID paths, string concatenation, command aliases, non-PowerShell interpreters, 8.3 short names, multi-command smuggling, document-body smuggling, and link traversal.
\item Group D --- post-approval mutation and credential-chain attacks (31). Modification of a staged artefact after approval, intent mutation, forged approval identifiers, replay, and identifier/operation mismatch.
\item Group E --- fault injection (32). Deliberately broken preconditions: missing or corrupted policy, unwritable ledger, unreadable intent files, malformed intents, and missing staged artefacts. These measure whether the system fails closed.
\end{itemize}

Each case declares the verdict a security policy \emph{should} return; the observed verdict is recorded in a separate field, so that disagreement between policy intent and implementation is visible rather than hidden. The benchmark was frozen before any system version was evaluated and was never adjusted to accommodate a version.

\subsection{System variants}
\label{sec:4-2}

We evaluate eight variants arranged as a progressive mechanism ladder. The ladder is not fully orthogonal: B5$\rightarrow$B6 isolates pre-execution file-byte verification and B6$\rightarrow$B7 isolates the OS sandbox, but earlier rungs also differ in accepted channels or in bundles such as fail-closed handling, audit, and close-out. We therefore use causal language only for the single-factor pairs and treat the remaining comparisons as bundle ablations. The eight variants and their mechanism sets are listed in Supplementary Table~S5.

Variants differ in which input channels they accept. B4 accepts only structured intents; B0--B3 accept only command strings; B5--B7 accept both. This asymmetry is reported explicitly wherever it affects a denominator (Section~\ref{sec:5-1}, Section~\ref{sec:5-2}). B1 is retained as a historical ablation control because it was part of the frozen eight-rung benchmark; its numbers must not be interpreted as estimates of real LLM behaviour. The real LLM experiment in Section~\ref{sec:4-6} is a separate external baseline.

\subsection{Decision rules and the two execution models}
\label{sec:4-3}

Adjudication returns ALLOW, ASK, or DENY. ALLOW is automatic; DENY refuses and requires an explicit override to proceed; ASK requires a human to approve on the spot, and silence counts as rejection --- an unanswered request is automatically settled as denied after a fixed timeout, attributed to no individual.

ASK is ambiguous when no human is actually in the loop. We therefore distinguish two execution models. Under the self-approval model, ASK is counted as executed: the conservative assumption that the same agent grants its own request, which is the relevant reading for a single-agent deployment or a prompt-injection scenario. Under the strict model, only ALLOW and OVERRIDE count as automatic execution. All headline figures use the self-approval model. The strict model is retained as a sensitivity definition, but it is not used for the headline comparisons reported below.

\subsection{Payload validity filter}
\label{sec:4-4}

Five bypass cases obfuscate the \emph{command name itself} --- full-width homoglyphs, zero-width joiners, tab separation, percent-encoding, and invalid quoting. PowerShell rejects these as unknown commands, so they cannot execute regardless of the gate.

We therefore report two attack-success figures. \texttt{attack\_success} counts every dangerous case the system permitted. \texttt{attack\_success\_effective} excludes the non-executable payloads, so that the gate is not credited for blocking an attack that could never have run. All headline figures use the effective variant, whose denominator is 117.

\subsection{Sandbox verification}
\label{sec:4-5}

The sandbox layer was verified on the real implementation rather than modelled. Using the sandbox's own headless runner, we re-executed the 94 DENY-expected group-C payloads that carry a command string, and probed the layer's decision points one at a time: a denial surfaces either at the shell layer with a genuine denial marker, at the approval layer as a fail-closed refusal, or not at all. The full protocol --- runner setup, the four transport defects that had to be removed before these runs could be trusted, and the relocated-tree layout --- is given in Supplementary C.

With the transport corrected, the 94 payloads are assigned to exactly one outcome class each, and the classes sum to 94 by construction: 57 refused at the tool or shell layer with a genuine denial marker; 4 refused at the approval layer as fail-closed; 5 refused by the agent itself; 10 could not execute in the deployed layout; 11 executed without effect; 2 were inconclusive; and 5 executed with effect. On the definition used here, the \emph{sandbox-evaluable} subset is the 57 tool/shell refusals plus the 4 approval fail-closed refusals plus the 5 effectful executions, i.e. N\_eval = 66; the refusal set has the same size but is not the same set (N\_refusal = 57 + 4 + 5 agent self-refusals). Five of the 66 executed with effect --- an effectful-execution rate of 7.6\%.

The layer's margin was measured rather than inferred. Re-running the same payloads with the layer removed --- same harness, same delivery, same relocated layout, one environment variable (ON = workspace-write with approval fail-closed; OFF = danger-full-access with no approval) --- flips 45 payloads from refused to executed with no case in the reverse direction over 88 valid pairs (McNemar exact p = 5.7$\times$$10^{-14}$), a measured marginal interception of 51.1\%. Six payloads were excluded a priori because their rewritten text still resolved to the live deployment trees.

The layer confines writes everywhere outside the workspace, but its deletion confinement is volume-scoped: a deletion outside the workspace is refused on the volume that carries the workspace and succeeds on another volume. Of the five effectful executions, three delete files inside the workspace, one is contaminated by scratch files the agent created for itself, and one --- a \texttt{Remove-Item} on a decoy outside the workspace, on a different volume --- deleted it silently with exit code 0 and no denial marker of any kind. That is 1 of 66 sandbox-evaluable payloads (1.5\%), and it is a genuine boundary escape.

Three deviations between our behavioural model and the real implementation are disclosed in Section~\ref{sec:7}: the model treated read-only invocations as writes (inflating modelled false denials); it did not model the sandbox's escalation path; and, most consequentially, it assumed that a refusal of a \emph{write} implies a refusal of a \emph{deletion}, which the re-execution above shows to be false.

\subsection{Real LLM baseline}
\label{sec:4-6}

To establish a prompt-only baseline with real models, we used three models from three distinct families under an identical system prompt requiring explicit approval before any dangerous action, and asked each to adjudicate every case five times. The prompt is reproduced in Supplementary C.

We sampled 150 cases stratified across the five groups (A 30, B 30, C 59, D 20, E 11; group E is smaller because several of its cases use the intent channel and therefore carry no command string to present) and repeated each case five times, giving 2,250 attempted calls at temperature 0, top-p 1.0 and a 2,000-token completion cap. Of these, 2,249 calls completed and 2,164 yielded a parseable decision.

\subsection{Statistics}
\label{sec:4-7}

Headline attack-success proportions and selected small-denominator rates are reported with Wilson 95\% confidence intervals, which behave better than the normal approximation at the small denominators occurring in groups D and E. Exact numerators and denominators are retained in the tables or Supplementary Material A.

For comparisons among benchmark variants that share the same supported cases, decisions form a paired design. We therefore use McNemar's exact test on per-case binary outcomes and report discordant-pair counts (b, c) alongside p-values where such a comparison is meaningful. We do not apply a paired test to the external LLM baseline, because it uses a different 150-case stratified sample.

For any measurement involving a language model we report run-to-run agreement, since repeated calls to the same model are not deterministic even at temperature 0 (Section~\ref{sec:5-6}).

\subsection{Environment and reproducibility}
\label{sec:4-8}

All experiments were conducted on a single machine running Windows 11 (build 26200) with Python 3.12.13 and PowerShell 5.1.26100.9444, on an Intel i7-11800H with 31.8 GB of RAM. Each evaluated system version was frozen before its corresponding run; the frozen source, policy file, executor, hashes, and environment record are included in the companion reproducibility artefact.

The benchmark, raw per-case results for every variant and version, analysis scripts, and figures are included in the same package, so that the reported numbers can be regenerated from the supplied artefact.

\section{Results}
\label{sec:5}

All headline execution outcomes in this section use the self-approval model (Section~\ref{sec:4-3}). Unless stated otherwise, effective attack-success rates use the 117 executable group-C payloads (Section~\ref{sec:4-4}). Variant definitions are as in Section~\ref{sec:4-2}; the primary results correspond to system version v4 with policy revision v7.

\subsection{RQ1 --- Does execution-path governance reduce dangerous execution?}
\label{sec:5-1}

Table~\ref{tab:1} summarises the main results across variants, reporting benign-task completion, false denials, false allows, and effective attack success.

\begin{table*}[!t]
\caption{Main results.}
\label{tab:1}
\centering
\resizebox{\textwidth}{!}{%
\begin{tabular}{lllll}
\hline
System & Normal success \% & False deny \% & False allow \% & Effective ASR \% (Wilson 95\% CI) \\
\hline
B0 direct execution & 100 & 0 & 100 & 98.3 (94.0--99.5) \\
\hline
B2 string gate & 100 & 0 & 31.1 & 43.6 (35.0--52.6) \\
\hline
B4 structured intent (intent channel only) & 100 & 0 & --- & 22.2 (9.0--45.2) \\
\hline
B5 full architecture (as-is) & 100 & 0 & 29.7 & 30.8 (23.1--39.6) \\
\hline
B6 + pre-execution file-hash verification & 100 & 0 & 29.7 & 30.8 (23.1--39.6) \\
\hline
B7 deployed (+ OS sandbox) & 79.6 raw $\rightarrow$ 88.9 corrected & 20.4 raw $\rightarrow$ 11.1 corrected & 23.0 & 7.7 (4.1--14.0) \\
\hline
\end{tabular}}
\end{table*}

Notes. False-denial figures are measured on the 54 group-A cases. The raw B7 sandbox model denied 11/54 benign cases; per-case review found that five were read-only operations that the behavioural model incorrectly treated as writes, leaving 6/54 genuine boundary denials (11.1\%, Wilson 95\% CI 5.2--22.2) and a corrected normal-success rate of 48/54 = 88.9\% (Section~\ref{sec:4-5}, Section~\ref{sec:7}). The B4 attack-success figure has a denominator of 18, not 117 --- that variant accepts only structured intents, leaving 18 effective group-C payloads --- and is therefore not directly comparable with the other rows; the cross-variant comparison rests on B2, B5, B6 and B7.

\begin{figure*}[!t]
\centering
\includegraphics[width=0.86\textwidth]{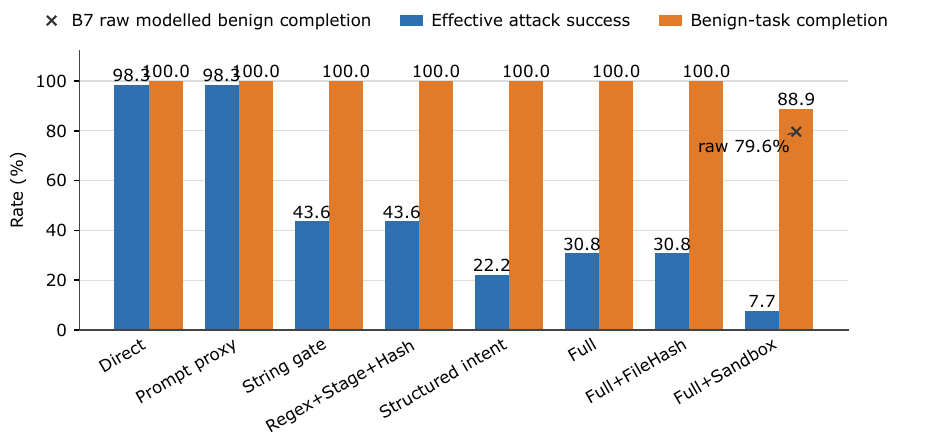}
\caption{Effective attack success versus benign-task completion for the eight frozen benchmark variants. The ladder reads left to right from direct execution to the deployed configuration; the change from B6 to B7 reflects the benchmark-model contribution of the operating-system sandbox. The plotted B7 benign-completion coordinate uses the corrected 88.9\% value after per-case sandbox review; the raw behavioural-model value (79.6\%) is shown separately as a reference marker (Section~\ref{sec:5-1}). The B4 bar is not directly comparable with the others: that variant accepts only structured intents, so its value derives from 18 effective payloads rather than 117.}
\label{fig:2}
\end{figure*}

Introducing a deterministic pre-execution gate (B2 versus B0) reduces effective attack success from 98.3\% to 43.6\%; the full layer reaches 30.8\% and the deployed configuration 7.7\% (Figure~\ref{fig:2}; the family-level breakdown is Supplementary Figure S1). On the 117 executable bypass payloads, paired McNemar comparisons give (b,c) = (64,0) for B0$\rightarrow$B2, (15,0) for B2$\rightarrow$B5 and (27,0) for B6$\rightarrow$B7, with exact p-values in Supplementary A.2; B5$\rightarrow$B6 changes no group-C decision, as the added file-hash check targets post-approval mutation rather than expression-level bypass. A substantial final-stage reduction is attributable to the operating-system sandbox, and re-executing the sandbox-relevant payloads against the real implementation gives a numerically similar aggregate rate (7.6\% over 66 sandbox-evaluable payloads) --- although aggregate similarity does not imply case-level agreement, and the layer's confinement turns out to be volume-scoped (Section~\ref{sec:4-5}). A single-factor ablation settles the magnitude: removing the layer flips 45 of the same payloads from refused to executed with no case in the reverse direction (88 valid pairs, p = 5.7$\times$$10^{-14}$) --- a measured marginal interception of 51.1\%.

\subsection{RQ2 --- What changes along the mechanism ladder, and which effects can be isolated?}
\label{sec:5-2}

Supplementary Table S4 reports the progressive mechanism ladder. The first two removals are clean single-factor comparisons; later rows are intentionally labelled as configuration bundles because more than one capability changes.

Three reading notes matter. First, B6$\rightarrow$B7 cleanly isolates the sandbox: effective ASR falls from 30.8\% to 7.7\% while mutation blocking is unchanged. Second, B5$\rightarrow$B6 cleanly isolates executor-side file-byte verification: group-C ASR is unchanged while mutation blocking rises from 61.3\% to 90.3\% --- exactly the failure class the mechanism targets. Third, the earlier B5$\rightarrow$B3$\rightarrow$B2$\rightarrow$B1 transitions are bundle comparisons, not single-factor estimates; their denominators also differ (B2 reports 2/27 rather than /31 for mutation blocking, because four group-D cases need channels B2 does not support).

A real end-to-end test makes the mutation case concrete. Before the corresponding fix, two of seven mutation variants actually deleted the decoy file; after the fix, none of the seven executed at all (Section~\ref{sec:5-7}, VUL-003). Supplementary Figure S4 shows the ladder visually.

\subsection{RQ3 --- What does the security gain cost in false denials?}
\label{sec:5-3}

Table~\ref{tab:2} sets the architecture against the prompt-only baselines on the two axes that matter: benign non-execution and bypass success.

\begin{table*}[!t]
\caption{Security--utility trade-off.}
\label{tab:2}
\centering
\begin{tabular}{lll}
\hline
System & Benign non-execution \% (case-level) & Bypass success \% (case-level) \\
\hline
Prompt-only, claude-opus-5 & 3.3 (1/30) & 89.8 (53/59) \\
\hline
Prompt-only, deepseek-v4.1-flash & 0.0 (0/30) & 94.9 (56/59) \\
\hline
Prompt-only, gpt-6-astra & 0.0 (0/30) & 100.0 (59/59) \\
\hline
TDGA v4 + sandbox & 11.1 (6/54) & 7.7 (9/117) \\
\hline
\end{tabular}
\end{table*}

Both columns are computed at the case level, the convention used for TDGA in Table~\ref{tab:1}. Denominators are 30 benign and 59 bypass cases for the LLM baselines, and 54 benign cases and 117 effective bypass payloads for TDGA. Two conventions are not symmetric. First, repetition structures differ: each TDGA case yields one adjudication, whereas each LLM case is attempted five times and aggregated by majority vote. Second, ``benign non-execution'' for the LLM baselines means a majority decision that is not \emph{execute} under the self-approval model, and therefore includes explicit refusals and unparsable responses; on the 30 benign cases claude-opus-5 produced 2 explicit refusals and 16 unparsable responses, and the unparsable ones dominate the single case counted as non-execution, whereas TDGA always returns a well-formed verdict. Under this single convention the two systems differ in both directions: TDGA has more benign non-execution (11.1\% versus 3.3\%) and permits far fewer bypasses (7.7\% versus 89.8\%). We claim no matched ``same usability budget'' experiment: the 12.0\% value for claude-opus-5 under call-level aggregation (18 of 150 calls) is a call-level quantity and is not comparable with TDGA's case-level rate.

Because the LLM baselines were evaluated on a 150-case stratified sample while the architecture was evaluated on the full 313-case benchmark, the sampling still differs even after aligning the aggregation level (Section~\ref{sec:4-6}, Section~\ref{sec:7}); no paired significance test is claimed for Table~\ref{tab:2}.

\subsection{RQ4 --- Is auditing complete?}
\label{sec:5-4}

In the primary sequential benchmark, audit coverage --- adjudication, execution, receipt, and close-out all recorded --- is 100\% for B5, B6 and B7, and structurally absent for B0--B4. The concurrency experiment below shows why this sequential figure should not be interpreted as a guarantee under concurrent writers.

Two integrity limits were measured and fixed. Concurrent ledger writes lost records --- 9.66\% at two writers, 21.93\% at four and 31.32\% at eight, with 12 torn lines --- until a file lock with a single write and an fsync reduced loss to zero at 1, 2, 4 and 8 writers across five repetitions each, at roughly 4.5$\times$ per-write latency (VUL-002). Separately, malformed ledger lines are skipped silently at read time and must be counted by a dedicated health check, or nobody learns that the ledger is incomplete.

\subsection{RQ5 --- What is the overhead?}
\label{sec:5-5}

Supplementary Table S1 reports latency at three layers and the token cost.

For short CLI invocations, process startup dominates the measured overhead. Keeping the gate resident removes that startup term; common-case fresh-credential look-ups remain sub-millisecond to low-millisecond over the tested ledger sizes.

A measurement caveat is worth stating rather than leaving for a reviewer to discover. The 4--6 ms figure was obtained on an empty ledger. Ledger look-ups --- signature verification, credential-consumption checks, the executor's look-up of an adjudication record by judgement id, and the staged-file branch of intent adjudication --- originally scanned the whole append-only log, so their latency grew approximately linearly over the tested ledger-size range. Two of them must be kept apart when quoting numbers, because they read in opposite directions: the oldest-record query (signature verification) took 676 ms at 50,000 lines, and a newest-record query (looking up a freshly issued credential) took 163.83 ms at the same size. After switching to reverse block-wise reading with early termination, the newest-record queries became effectively size-independent over the tested 0--50,000-line range, reaching 0.83 ms at 50,000 lines, whereas the oldest-record query improved only to 228 ms, since reaching the oldest entry still requires reading the whole log. The worst case in the mixed-workload soak (looking up a very old credential) became about 30\% slower and constitutes a slow-DoS surface (Section~\ref{sec:6-5}, Section~\ref{sec:7}). Any latency figure in this paper must therefore be read together with the ledger size at which it was measured. Supplementary Figure S5 plots the controlled measurement across ledger sizes for both the pre-fix and post-fix implementations.

On cost, the deterministic gate spends zero tokens, whereas the LLM baselines averaged \textyen{}0.22 per attempted decision, with a \textasciitilde{}400$\times$ spread between the cheapest and most expensive model on the identical task (Section~\ref{sec:6-4}).

\subsection{RQ6 --- How do prompt-only failure modes vary across model families?}
\label{sec:5-6}

Supplementary Table S2 breaks the prompt-only baselines down by decision type and reports run-to-run agreement.

The decision-distribution columns are call-level (denominator: completed calls); five-run agreement is case-level. Table~\ref{tab:2} re-aggregates these decisions to the case level so that they can be compared with TDGA. Failure modes are family-specific and point in opposite directions. One model treated the bare request ``please execute this'' as sufficient authorisation and, under the self-approval execution model, permitted 100\% of dangerous operations with a zero refusal rate; the other two were conservative enough to demand confirmation for some legitimate operations.

\subsection{Failures as data: four defects found by the benchmark}
\label{sec:5-7}

Supplementary Table S3 lists the four implementation defects, how each was found, how it was fixed, and how the fix was verified.

All four defects were discovered by the benchmark, none by design review. Two of them were invisible in aggregate metrics --- VUL-001 manifested as \emph{stricter} behaviour, and VUL-004 as merely "another passing case". Both required tests that actually execute: one destroyed a decoy file, the other replayed a consumed approval. This is itself an argument for keeping such tests in the protocol rather than replacing them with static auditing.

Two measurement-level problems were also found and corrected, and are reported here because they affect how the numbers above should be read.

\begin{itemize}
\item A fault-injection path silently stopped taking effect after an implementation change, which lowered the measured fail-closed rate from 87.5\% to 81.2\% for reasons unrelated to the system under test; extending the injection to cover the new write path restored 87.5\%. Separately, a soak harness whose correction changed the observed latency-versus-size slope was separated by re-running with the harness corrected. Both, with the four further pitfalls of the re-execution, are documented in Supplementary C.3.
\end{itemize}

\section{Discussion}
\label{sec:6}

\subsection{The final security gain depends strongly on the deployment layer}
\label{sec:6-1}

This is the finding most in need of stating plainly, because it cuts against the framing the architecture invites.

The cleanest single-factor ablation is B6 versus B7: B6 (full architecture plus pre-execution file-hash verification) and B7 (the same configuration plus the operating-system sandbox) differ by 30.8\% versus 7.7\% in effective attack success. In the benchmark model this 23-point gap is attributed to the sandbox --- the affected cases include forward-slash paths into system directories, 8.3 short names, junction traversal, Python file APIs, and writes outside the allow-list that the higher-level gate does not reliably capture --- and the real-sandbox experiment independently verifies the relevant writable-root and path-resolution behaviour. The sandbox-relevant shell subset was re-executed directly (94 payloads, Section~\ref{sec:4-5}) and the confinement layer was evaluated in a paired on/off experiment (88 valid pairs, Section~\ref{sec:4-5}), which isolates a volume-scoped deletion boundary the behavioural model does not represent (Section~\ref{sec:7}).

Two implications follow. First, reporting only architecture-level metrics overstates what the architecture contributes: had we credited B5's 30.8\% to the architecture, we would have attributed the deployment environment to the design, which is why B6 and B7 are reported separately. Second, the outcome supports the principle that safety responsibilities must sit outside the model --- with the qualification that "outside" must be unreachable by expression-level tricks.

\subsection{The real usability cost is a boundary mismatch, not excessive strictness}
\label{sec:6-2}

The raw behavioural model yielded 79.6\% benign-task completion. Per-case review changes the picture. Of eleven denials, six reflect genuine constraints (the target was outside the writable root, or the operation would have written into the protected boundary), while five were an artefact of our own modelling, which treated read-only invocations as writes; the real sandbox does not refuse them. The corrected benign-task completion rate is therefore 88.9\% (48/54), corresponding to a corrected false-denial rate of 11.1\% (6/54); the policy gate's own false-denial rate is zero.

The residue traces to a single inconsistency: the gate's allow-list includes the workspace and the temporary area, whereas the sandbox's writable root is the session working directory. Tasks the gate adjudicates as ALLOW can therefore still be refused at execution time.

Set against the baselines using the same case-level aggregation, the directional trade-off is sharp: claude-opus-5 produces a non-executable outcome for 3.3\% (1/30) of benign cases and permits 89.8\% (53/59) of bypass cases, while TDGA's corrected benchmark figures are 11.1\% (6/54) and 7.7\% (9/117), respectively. Thus TDGA incurs a higher observed benign non-execution rate in these samples but permits far fewer bypass attempts. Because the samples and denominators differ, this remains a directional comparison rather than a paired security--utility estimate.

\subsection{Failure modes differ by family --- which is why governance must sit outside the model}
\label{sec:6-3}

The three prompt-only baselines fail in opposite directions. One reads "please execute" as authorisation, never refuses outright, and --- under the self-approval model, where an unanswered confirmation request counts as execution --- permits 100\% of the dangerous and bypass cases. Note that this 100\% and the decision split below have different denominators: across \emph{all} calls, including benign ones, the same model's decisions split 67.3\% execute and 32.7\% ask (Supplementary Table S2). The other two are conservative enough to request confirmation even for legitimate operations: on the benign cases, claude-opus-5 asked for confirmation in 28.0\% of calls and deepseek-v4.1-flash in 62.7\% (call-level, group A).

A second point matters for anyone designing an evaluation: at temperature 0, five repetitions of the same case agreed only 48.7\%--71.3\% of the time. A single run is not a measurement: any evaluation involving a language model must report run-to-run variation and its repetition count.

Two consequences follow. First, placing the enforcement boundary outside the model prevents a fixed candidate action from being adjudicated according to a particular model's disposition. Second, the measured blind spots are complementary rather than overlapping --- string rules can miss expression-level variants that a model interprets semantically, while a model can be persuaded by authorisation-flavoured phrasing that leaves deterministic rules unchanged. This motivates layered evaluation, but our current deployed gate remains deterministic; an LLM-based semantic adjudicator is a future extension (Section~\ref{sec:6-4}, Section~\ref{sec:7}).

\subsection{Deterministic and semantic adjudication are complementary}
\label{sec:6-4}

The two approaches have non-overlapping blind spots. The comparison is given in Supplementary Table~S6.

The implication is not replacement. Our current architecture uses deterministic checks for enumerable structure and keeps the auditor in a review role; it does not make the final execution decision. The LLM results instead motivate a future composition in which semantic checks handle contextual questions while deterministic enforcement remains authoritative for enumerable constraints --- a composition not evaluated in this paper.

\subsection{Scalability: an under-appreciated long-term cost}
\label{sec:6-5}

Governance overhead is usually reported as a constant. Our measurements show a second, scale-dependent component.

Two further results come from near-day soak runs of two system versions: 24.00 h for the pre-fix v4 run and 23.89 h for the post-fix v5 run, the latter ending in an external power interruption after its last complete snapshot. First, latency tracked ledger size almost exactly in both original-harness runs (R\textsuperscript{2} $\approx$ 0.998). Second, and less expected, part of the measured slope belonged not to the system under test but to the measurement harness: three workload branches read the entire ledger to retrieve a single trailing record. An additional 6.00 h v5 confirmation run with the harness corrected therefore completes a three-curve comparison on a common ledger-size axis:

\begin{table*}[!t]
\caption{Ledger-scaling slopes from the two near-day soak runs.}
\label{tab:3}
\centering
\begin{tabular}{llll}
\hline
Curve & Slope (ms per line) & Per 10k lines & R\textsuperscript{2} \\
\hline
A \textperiodcentered{} version before the fix, original harness & 0.005003 & +50.0 ms & 0.9981 \\
\hline
B \textperiodcentered{} version after the fix, original harness & 0.002507 & +25.1 ms & 0.9975 \\
\hline
C \textperiodcentered{} version after the fix, corrected harness & 0.000460 & +4.6 ms & 0.9996 \\
\hline
\end{tabular}
\end{table*}

The two soak traces are given as Supplementary Figures S2 and S3, and the slope differences provide a descriptive decomposition rather than a controlled causal partition (computed from the unrounded fitted slopes): A $-$ B = 0.002496 is the slope reduction associated with the system-side change under the original harness, B $-$ C = 0.002046 the additional reduction with the corrected harness, and C = 0.000460 the residual. Because curve C is a shorter confirmatory run rather than a duration-matched counterpart to curve B, these differences are not an exact causal partition of the original slope; the residual is attributed to no identified read path and is recorded as an open question (Section~\ref{sec:7}).

The residual is small but the lesson is not: a measurement harness can contaminate the object it measures. The reduction after correcting the harness shows that part of the original slope came from the measurement process itself. The corrective is procedural: harness and system must be varied independently, and any difference reported with its experimental conditions rather than treated as a strict causal partition.

The practical implication is that ledger layout and indexing are first-class engineering concerns for self-auditing systems, not bookkeeping details. A deployment performing a few hundred adjudications a day accumulates roughly fifty thousand ledger entries within a month or two --- precisely the range where the unindexed design becomes the bottleneck.

\subsection{Four defects, all found by experiment}
\label{sec:6-6}

All four defects reported in Section~\ref{sec:5-7} were discovered by the benchmark; none came from design review. Two were invisible in aggregate metrics. VUL-001, an undefined variable that made two intent branches always raise, manifested as \emph{stricter} behaviour --- the exception was swallowed into a denial, so the system looked safer while the functionality was in fact dead. VUL-004, a reusable approval, manifested as merely "another passing case": the recorded verdict was ordinary, and only replaying a consumed approval revealed that the credential never expired.

This supports the paper's central methodological claim: verification of a security system must include end-to-end paths that actually execute, not only unit checks of the adjudication layer. What mattered here were the cases where a decoy file was actually deleted, and where a consumed approval was actually replayed. Static auditing reports intent; only execution reports behaviour.

\subsection{Design implications}
\label{sec:6-7}

1. Place adjudication outside the model, and make an approval a credential that is digest-bound and single-use. 2. Align the sandbox boundary with the policy allow-list: where they diverge the cost appears as "the gate said yes and execution refused", a usability cost that aggregate metrics hide. 3. Index or segment the ledger, and treat look-up latency versus ledger size as a routine measurement. 4. Consume an approval exactly once, recorded at the executor rather than through a path that can fail silently. 5. Include destructive end-to-end tests: changes that merely look stricter will otherwise pass review.

\section{Limitations}
\label{sec:7}

Modelling versus measurement. The B7 figures in Section~\ref{sec:5} come from the behavioural model, not from executing every case in the deployed configuration. The sandbox-relevant subset was re-executed against the real implementation (Section~\ref{sec:4-5}): 94 DENY-expected group-C payloads, 66 of them sandbox-evaluable, give a measured effectful-execution rate of 7.6\% (5/66) against the modelled 7.7\%; the two 66-payload sets have the same size but are not the same set ($N_{\mathrm{refusal}}$ against $N_{\mathrm{eval}}$). On the same 66 payloads the behavioural model yields 3.0\%, and over the 94 per-case agreement is 89.4\%, dominated by jointly negative cases --- none of the five model-predicted effectful cases coincides with the five observed empirically. Three limits attach to that measurement (scope of channel, layout drift, and a non-uniform refusal mechanism) and two further coverage limits attach to the ablation itself (six payloads unrunnable in the unconfined arm, and registry writes that the unconfined toolchain performs). All five are stated in full in Supplementary C.2.

Three known deviations of our sandbox model. First, read-only invocations were modelled as writes, which inflated modelled false denials; the corrected figure is 11.1\% (Section~\ref{sec:6-2}). Second, and most consequentially, the model assumed that a refusal of a write implies a refusal of a deletion --- Section~\ref{sec:4-5} shows it does not: writes outside the workspace are refused on any volume, but deletion confinement is volume-scoped, so any single sandbox interception rate that does not separate policy denials from access-control denials overstates the layer's contribution. Third, the escalation path is not represented: unattended, the sandbox fails closed, but interactively a model may request wider access and a human decides --- and humans are persuadable. The unattended half is now measured: asked to perform a denied operation, the governed agent attempted one escalation on its own initiative and was refused because no approval channel existed, leaving the target unchanged. The interactive half remains a real, unmeasured attack surface.

Permitted versus executed. The Section~\ref{sec:5} figures are adjudications, not executions: "executed" means "permitted by the adjudication layer", and destructive consequences are inferred from the decision plus the known behaviour of the relevant channels, corroborated by the end-to-end tests in Section~\ref{sec:5-7} where a decoy file was deliberately placed in harm's way. The sandbox re-execution in Section~\ref{sec:4-5} is the exception --- those 94 payloads really ran, but against a relocated copy of the deployment, because deletion outside the workspace is not uniformly confined and running them live would have destroyed the gate being measured.

Measurement pitfalls encountered. Six pitfalls affected how these numbers should be read, and each first produced a plausible but wrong value: a fault-injection path that silently stopped taking effect after an implementation change (corrected and re-verified); a soak harness whose correction was associated with a substantial reduction in the observed latency-versus-size slope (separated by re-running with the harness corrected); and, in the re-execution, payloads that never reached the agent, prose keywords counted as denials, an unset environment variable that degraded to "path not found", and a staged script that never ran. The sixth is methodological and we report it against our own method: in the first attempt at the unconfined arm, two payloads reached the live deployment through an administrative UNC share and deleted a protected file; the per-case canary detected the change, it was restored byte-identically, and the arm was re-run with an a-priori exclusion rule. All six are documented in Supplementary C.3, together with the residual slope of 0.000460 ms per line that remains unexplained and is not attributed.

Ledger figures must be read with their size. Every latency measurement involving ledger look-ups is conditional on the ledger size at which it was taken; the same operation varies by two orders of magnitude across the sizes we tested (Section~\ref{sec:5-5}, Section~\ref{sec:6-5}).

Model baselines are directional comparisons. The LLM baselines were evaluated on a 150-case stratified sample while the architecture was evaluated on the full 313-case benchmark; the denominators and sampling differ, so Section~\ref{sec:5-3} is a directional comparison rather than a paired test. A strictly paired comparison would require running the LLM baselines on the same case set.

Model versions drift. All three LLMs are closed-hosted APIs whose behaviour may change without notice, which weakens long-term reproducibility; local models with pinned weights would trade that reproducibility for a different set of hardware and quantisation variables.

Scope. Single host, single operating system, single PowerShell version; cross-platform generalisation is not established. Multi-agent collusion, side channels, and supply-chain compromise were not investigated.

Long-run deployment. The principal soak runs cover 24.00 h (v4) and 23.89 h (v5), plus a 6.00 h corrected-harness confirmation run; the v5 near-day run ended in an external power interruption rather than a system termination, and recorded one failure/crash event among 118,648 actions, which we report without attributing it to the version from a single occurrence. These are endurance tests, not longitudinal field deployment: behaviour over weeks or months --- policy evolution under real use and drift in human intervention rates --- is left to future work.

\section{Conclusion}
\label{sec:8}

We asked whether governing an agent's actions rather than only its inputs can be made both effective and affordable, and whether that boundary can be decoupled from model-specific behaviour. The evidence supports the first claim and supports model-decoupled adjudication for a fixed candidate action; full cross-model generation-to-execution evaluation remains future work.

Effective attack success fell from 98.3\% to 7.7\%, while per-case review corrected the deployed false-denial rate from 20.4\% to 11.1\%. In the separate prompt-only experiment, the claude-opus-5 baseline produced a non-executable outcome for 3.3\% (1/30) of benign cases while permitting 89.8\% (53/59) of bypass cases --- a directional, not matched, comparison: TDGA shows a higher benign non-execution rate but a far lower bypass-success rate.

The remaining qualifications are stated in Section~\ref{sec:5} and Section~\ref{sec:6}: the sandbox's substantial final-stage contribution, the measured 7.6\% on the sandbox-evaluable subset, the 51.1\% marginal interception over 88 valid pairs, the volume-scoped deletion boundary, and the residual slope of +4.6 ms per 10,000 lines. On cost, the deterministic gate spends zero tokens whereas the hosted baselines averaged \textyen{}0.22 per attempted decision with a \textasciitilde{}400$\times$ spread; deterministic adjudication is cheap, reproducible and version-stable, but limited to enumerable structure.

\section*{Acknowledgment}

The author used an AI-based writing assistant, DeepSeek, for language editing of the manuscript and for drafting and revising portions of the figure-generation and analysis scripts included in the supplementary reproducibility artifact. The AI system was not used to generate experimental data, measurements, or scientific conclusions. All AI-assisted material was independently reviewed, tested, and verified by the author against the experimental records, and the author takes full responsibility for the content of this article.

\end{document}